\documentclass[prb]{revtex4}
\usepackage[dvips]{color}
\usepackage{graphicx}
\usepackage{dcolumn}
\usepackage{bm}

\begin{document}


\title {Quantized spin Hall effect in $^3$He-A and other 
$p$-wave paired Fermi systems}

\author{Jun Goryo}
 \affiliation{
 Dept. of Phys. and Math., Aoyama Gakuin University, 5-10-1, Fuchinobe, 229-8558, Japan}

\author{Mahito Kohmoto}
\affiliation{ Institute for Solid State Physics, University of
Tokyo\\ 5-1-5 Kashiwanoha, Kashiwa, Chiba, 277-8581, Japan}

\author{Yong-Shi Wu}
\affiliation{
Department of Physics, University of Utah,
Salt Lake City, Utah 84112}


\date{\today}

\begin{abstract}
In this paper we propose the quantized spin Hall effect (SHE) in
the vortex state of a rotating $p$-wave paired Fermi system in an
inhomogeneous magnetic field and in a weak periodic potential. It
is the three dimensional extension of the spin Hall effect for a
$^3$He-A superfluid film studied in Ref. \cite{G-K}. It may also
be considered as a generalization of the 3D quantized charge Hall
effect of Bloch electrons in Ref. \cite{K-H-W} to the spin
transport. The A-phase of $^3$He or, more generally, the $p$-wave
paired phase of a cold Fermi atomic gas, under suitable conditions
should be a good candidate to observe the SHE, because the system
has a conserved spin current (with no spin-orbit couplings).

\end{abstract}

\pacs{Valid PACS appear here}
\maketitle

\section{Introduction}

Recently, spin transport in condensed matter systems has received
wide attention. One of the most striking phenomena is the spin Hall
effect (SHE) in semiconductors \cite{M-N-Z}. In this effect, the
spin current is driven by an electric field, and flows in a
direction perpendicular to it. Time reversal symmetry is respected,
the situation is remarkably different from the Hall effect of
charged currents in a magnetic field. However, the SHE in
semiconductors needs appreciable spin-orbit couplings, which make
the spin current non-conserved. This causes several complications in
theoretical study, and makes the relationship between the spin
current and the experimentally observable spin accumulation
intricate and obscure.

In the literature, another class of condensed matter systems,
namely superfluids, have been suggested in which the SHE may be
observable without the need of spin-orbit couplings. It has been
shown that in the vortex state of two-dimensional (2D) $d$-wave
superconductors, the spin current can be generated by an
inhomogeneous magnetic field and flows perpendicular to the field
gradient \cite{V-V}. Moreover, a similar spin Hall effect was
suggested for a rotating superfluid $^3$He film in a non-uniform
magnetic field \cite{G-K}. In this case, time reversal symmetry is
broken. If the quasiparticle spectrum has a gap and the Fermi
level lies within the gap, the spin Hall conductance can be
expressed as a topological number (the first Chern number) and
therefore is quantized, in the units of $\mu_{\rm He} /8\pi$ with
$\hbar=c=1$ where $\mu_{\rm He}$ is the Zeeman coupling constant
for $^3$He atom. This effect is thus the spin analogue of the
quantized charge Hall effect of Bloch electrons
\cite{TKNN,Kohmoto}. The distinctive feature of the SHE in
superfluids is the absence of spin-orbit couplings. In general,
spin-orbit couplings exist in superconductors or semiconductors,
because of the presence of the internal crystal fields or
confining fields (in the case of heterostructures). But spin-orbit
couplings do not exist in superfluid $^3$He, of which the
constituents are neutral atoms. (Recall that spin-orbit couplings
exist only for charged particles). Superfluid $^3$He is in the
spin-triplet $p$-wave state \cite{He} and the spin rotational
symmetry is broken spontaneously. However, one of $^3$He
superfluid phases, i.e. the $^3$He-A phase, has a residual
rotational symmetry around the $z$-axis in the spin space. In this
phase the system has a conserved spin current \cite{He, G-K}.
Therefore, $^3$He-A film should be a good candidate to observe the
QSHE without worrying about complications due to non-conservation
of the spin current. This motivates us to study this system in
more general three dimensional cases.

In this paper, we propose the quantized SHE in the vortex state of
a rotating $p$-wave paired Fermi system in a non-uniform magnetic
field and in a periodic potential in three dimensions. Our
proposal is a generalization of the SHE for a thin film of
$^3$He-A superfluid studied in ref. \cite{G-K}. On the other hand,
it is the spin analogue of the three-dimensional quantum charge
Hall effect (3DQHE) for Bloch electrons suggested in ref.
\cite{K-H-W}. Our discussions are valid also for the vortex state
of a cold Fermi atomic gas that is in the $p$-wave pairing regime
\cite{note}. 

\section{Vortex state in the $^3$He-A phase}

We use the same notations as in ref. \cite{G-K}. The
gap matrix for a $^3$He superfluid is defined as
$\hat{\Delta}_{\alpha\beta}({\bf x},{\bf y}) =\sum_{\gamma\delta}
V_{\alpha\beta\gamma\delta}({\bf x},{\bf y}) \langle
\psi_\gamma({\bf x})\psi_\delta({\bf y})\rangle$, where
$\psi_\alpha({\bf x})$ is the Fermion field for the $^3$He atom with
spin $\alpha=\uparrow,\downarrow$,$\langle...\rangle$ the thermal
expectation value, and $V_{\alpha\beta\gamma\delta}({\bf x},{\bf
y})$ the pairing interaction. Under the U(1) gauge transformation
$\psi_{\alpha}({\bf x}) \rightarrow e^{-i\theta ({\bf
x})}\psi_{\alpha}({\bf x})$, the gap matrix is transformed as
$\hat{\Delta}_{\alpha\beta}({\bf x},{\bf y})\rightarrow
\hat{\Delta}_{\alpha\beta}({\bf x},{\bf y})e^{-i \theta({\bf x}) - i
\theta({\bf y})}$. The matrix can be written in a form $
\hat{\Delta}_{\alpha\beta}({\bf x},{\bf y}) = \Phi\left({\bf
R}_{G}\right) \int \frac{d^3k}{(2 \pi)^3}
\hat{\Psi}_{\alpha\beta}({\bf k}) e^{i {\bf k} \cdot({\bf x}- {\bf
y})} , $ where $\Phi\left({\bf R}_{G}\right)$ denotes the gap
amplitude depending on the center of mass of a Cooper pair ${\bf
R}_{G}=({\bf x} + {\bf y})/2$ and $\hat{\Psi}_{\alpha\beta}({\bf
k})$ the order parameter. For the spin triplet pairing state, the
order parameter is written by using ${\bf d}({\bf k})$-vector in the
spin space\cite{He} as $\hat{\Psi}_{\alpha\beta}({\bf k})=i {\bf
d}({\bf k}) \cdot ({\bf {\sigma}} \sigma_2)_{\alpha\beta}$, where
${\bf \sigma}=(\sigma_1,\sigma_2,\sigma_3)$ is the Pauli matrix.

In the $^3$He-A phase, the ${\bf d}$-vector is given by
\begin{eqnarray}
{\bf d}({\bf k})={\bf e}_z (k_x + i k_y).
\label{d-vector}
\end{eqnarray}
Clearly the order parameter is invariant under rotations around the
$z$-axis in the spin space. As we mentioned, the presence of the
spin rotational symmetry is crucial for defining a conserved spin
current. In the particle-hole representation, the quasiparticles are
described by a two-component field
$$
\Psi({\bf x})=
\left(\begin{array}{l}
\psi_{\uparrow}({\bf x}) \\
\psi_{\downarrow}^{*}({\bf x})
\end{array}
\right).
$$
The Hamiltonian for a rotating superfluid with angular velocity
${\bf \Omega}$ in the  rotating frame reads \cite{G-K}
\begin{eqnarray}
H&=&\int d^3x d^3y \Psi^{\dagger}({\bf x}) {\cal{H}}(\hat{\bf p} ,
{\bf x}, {\bf y})\Psi({\bf y})
\end{eqnarray}
with
\begin{eqnarray}
{\cal {H}}(\hat{\bf p}, {\bf x}, {\bf y})&=&
\left(
\begin{array}{cc}
\epsilon(\hat{\bf p} - m {\bf R}) \delta^3({\bf x}-{\bf y})&
{\Delta}^{\rm inv}_{\rm A}({\bf x},{\bf y}) e^{-\frac{i}{2}(\varphi({\bf x})
+\varphi({\bf y}))} \nonumber\\
-{\Delta}_{\rm A}^{\rm inv*}({\bf x}, {\bf y})e^{\frac{i}{2}(\varphi({\bf x})
+\varphi({\bf y}))} & - \epsilon(\hat{\bf p} + m {\bf R})
\delta^3({\bf x}-{\bf y}) ,
\end{array}
\right)
\nonumber\\
\epsilon(\hat{\bf p}\pm m {\bf R})&=&\frac{1}{2m}(\hat{\bf p} \pm m
{\bf R})^2 - \mu_{\rm F} ,
\nonumber\\
\hat{\bf p}&=&-i {\bf {\nabla}},
\nonumber\\
{\bf R}&=&{\bf \Omega} \times {\bf x} ,
\nonumber
\end{eqnarray}
where ${\Delta}^{\rm inv}_{\rm A}({\bf x}, {\bf y})$ is
the gauge invariant part of the spin-traced-out gap function for
A-phase $\Delta_{\rm A}({\bf x},{\bf y})\equiv (1/2) {\rm
Tr}[\hat{\Delta}({\bf x},{\bf y}) \sigma_1]$ and $(\varphi({\bf
x})+\varphi({\bf y}))/2$ the gauge dependent phase. Namely, we have
a relation $\Delta_{\rm A}({\bf x}, {\bf y})={\Delta}^{\rm inv}_{\rm
A}({\bf x}, {\bf y}) \exp[-i (\varphi({\bf x})+\varphi({\bf
y}))/2]$. Moreover, since ${\Delta}^{\rm inv}_{\rm A}({\bf x}, {\bf y})$
depends only on the relative coordinate ${\bf x}-{\bf y}$, for $p$-wave 
pairing we have ${\Delta}^{\rm inv*}_{\rm A}({\bf y}, {\bf x})= 
-{\Delta}^{\rm inv*}_{\rm A}({\bf x}, {\bf y})$. From the above equations 
the correspondence between the case of a superconductor in an applied 
magnetic field and that of a rotating superfluid is clear, as shown in 
Table \ref{table1}. From the correspondence, one naturally expects that 
a vortex state emerges when $\Omega$ is larger than a certain critical 
value $\Omega_{c1}$.
\begin{table}
\caption{The correspondence between a superconductor in an applied
magnetic field and a rotating superfluid.}
\begin{center}
\begin{tabular}{c|c}
superconductors &  the rotating superfluid
\\ \hline
~ $e$ & $m$ \\
${\bf A}=\frac{1}{2} {\bf B} \times {\bf r}$ &
 ${\bf R}\equiv {\bf \Omega} \times {\bf r}$  \\
${\bf B}$ & $2 {\bf \Omega}$
\end{tabular}
\end{center}
\label{table1}
\end{table}
The vortex lines are parallel to ${\bf \Omega}$, and the
vortices form a lattice on the plane perpendicular to $\hat{\bf
n}={\bf \Omega}/\Omega$. The center ${\bf x}^i$ of the $i$-th vortex
can be expressed in terms of the primitive vectors ${\bf a}$ and
${\bf b}$ of the lattice as
$$
{\bf x}^i = l_i {\bf a} + n_i {\bf b} + z \hat{\bf n} ,
(l_i,n_i\; {\rm integers}; z~{\rm real}),
$$
where the phase $\varphi({\bf x})$ of the gap function satisfies
\begin{eqnarray}
\nabla \times \nabla \varphi ({\bf x}) &=&
2 \pi  \hat{\bf n} \sum_i \delta^3({\bf x} - {\bf x}^i).
\label{winding}
\end{eqnarray}
Since a Cooper pair has a "charge" $2m$, a vortex has a "flux"
$\pi / m$, and one obtains
$$
2 \Omega=\frac{\pi}{m |{\bf a}\times{\bf b}|}.
$$
This relation is consistent with the correspondence between the
$\Omega$-flux in the unit cell of the vortex lattice and the flux
quanta $\pi/e$ in superconductors which  is a half of the unit
flux $2\pi/e$. The difference is due to the fact that the charge
of a Cooper pair in superconductors is $2e$.

Now we proceed to examine the lattice symmetry of this vortex
state. We will see that the system has a "magnetic" translational
symmetry with respect to {\it two cells} of the vortex lattice,
since a vortex has a "flux" $\pi / m$, which is a half of the
"unit flux" $2\pi / m$. In an analogy to the system of charge
particles on a lattice, we consider the "magnetic" translation
operator \cite{G-K}
$$
T_{\delta {\bf r}}=\exp[i \delta {\bf r} \cdot (\hat{\bf p}
+ \tau_3 m {\bf R})],
$$
where $\tau_3$ is the third component of the Pauli matrices in
particle-hole space. By using the gauge degrees of freedom, one can
introduce the following constraints without conflict with Eq.
(\ref{winding})\cite{G-K}
\begin{eqnarray*}
\varphi({\bf x}+{\bf a})&=&\varphi ({\bf x})- m {\bf a}\cdot{\bf R},
\nonumber\\
\varphi({\bf x}+ {\bf b})&=&\varphi ({\bf x}) - m {\bf b}\cdot{\bf R}.
\end{eqnarray*}
Then we see that the system has the translational symmetry
generated by $T_{{\bf a}}$ and $T_{2 {\bf b}}$
\begin{eqnarray*}
[H,T_{{\bf a}}]=[H,T_{2 {\bf b}}]=0,
[T_{{\bf a}},T_{2 {\bf b}}]=0.
\end{eqnarray*}

Let us suppose that there is a periodic potential, e.g. as what
happens in an optical lattice
\begin{equation}
U({\bf x})=U_0 \cos {\bf K} \cdot {\bf x} \, .
\label{potential}
\end{equation}
Then the Hamiltonian density becomes
\begin{equation}
{\cal {H}}(\hat{\bf p},{\bf x},{\bf y}) \rightarrow
{\cal {H}}_U (\hat{\bf p},{\bf x},{\bf y})=
{\cal{H}} (\hat{\bf p},{\bf x},{\bf y})
+ U({\bf x}) \tau_3 \delta^3({\bf x}-{\bf y}).
\label{Hu}
\end{equation}
Suppose that ${\bf K}$ is {\it not} in the $ab$-plane, i.e. ${\bf
K}\cdot\hat{\bf n}\neq0$ and consider a cell spanned by the
following vectors
\begin{eqnarray}
{\bf a}^{\prime}&=& {\bf  a} + \alpha_n \hat{\bf n},
\nonumber\\
2 {\bf b}^{\prime}&=& 2 ({\bf  b} +  \beta_n \hat{\bf n}),
\nonumber\\
{\bf c}^{\prime}&=&\frac{2 \pi}{{\bf K}\cdot\hat{\bf n}}\hat{\bf n},
\end{eqnarray}
where
\begin{eqnarray*}
\alpha_n&=&\frac{1}{{\bf K}\cdot\hat{\bf n}} (2 \pi - {\bf K}\cdot{\bf a}),
\nonumber\\
\beta_n&=&\frac{1}{{\bf K}\cdot\hat{\bf n}} (2 \pi - {\bf K}\cdot{\bf b}).
\end{eqnarray*}
We see that our system has a three dimensional periodicity
represented by the commutation relations
\begin{equation}
[H, T_{{\bf a}^{\prime}}]=[H,T_{2{\bf b}^{\prime}}]=[H,t_{{\bf c}^{\prime}}]=0,
\label{symmetry}
\end{equation}
where
$
t_{{\bf c}^{\prime}}=\exp[i{\bf c}^{\prime} \cdot{\hat{\bf p}}]
$
is the usual translational operator.
We also note that
\begin{eqnarray}
[T_{{\bf a}^{\prime}}, T_{2 {\bf b}^{\prime}}]
=[T_{2{\bf b}^{\prime}}, t_{ {\bf c}^{\prime}}]=
[t_{{\bf c}^{\prime}}, T_{ {\bf a}^{\prime}}]=0.
\label{algebra}
\end{eqnarray}

The Bogoliubov-de Gennes equation for our system reads
\begin{eqnarray}
\int d^3y {\cal{H}}_U(\hat{\bf p},{\bf x},{\bf y}) \Phi_E({\bf
y})&=&E\Phi_E ({\bf x}),
\nonumber\\
\Phi_E({\bf x})&=&
\left(
\begin{array}{c}
U_E ({\bf x})
\\
 -V_E^*({\bf x})
\end{array}
\right) \, ,
\label{BdG}
\end{eqnarray}
where $(U_E({\bf x}), V_E({\bf x}))$ is related to the fermion
operators as
$$
\left(\begin{array}{l}
\psi_{\uparrow}({\bf x})
\\
\psi_{\downarrow}^*({\bf x})
\end{array}
\right)
=\sum_E
\left(
\begin{array}{cc}
U_E({\bf x}) & V_E({\bf x})
\\
-V_E^*({\bf x}) & U_E^*({\bf x})
\end{array}
\right)
\left(
\begin{array}{l}
\gamma_{E\uparrow}
\\
\gamma_{E\downarrow}^*
\end{array}
\right) \, ,
$$
and $\gamma_{E\alpha}^*$ and $\gamma_{E\alpha}$ are the creation and
annihilation operator, respectively, of the Bogoliubov
quasiparticles. In accordance with Eqs. (\ref{symmetry}) and
(\ref{algebra}), the solution of Eq. (\ref{BdG}) is of the Bloch
form
\begin{eqnarray*}
\Phi_{n{\bf k}}({\bf x})&=&e^{i {\bf k} \cdot {\bf x}}
u_{n {\bf k}}({\bf x}),
\nonumber\\
E&=&E_{n{\bf k}},
\end{eqnarray*}
where $n$ is the band index and ${\bf k}$ is the crystal momentum
defined in the 'magnetic' Brillouin Zone
$$
\left\{
\begin{array}{l}
-\pi / a^{\prime} \leq k_x < \pi / a^{\prime},
\\
-\pi / 2 b^{\prime} \leq k_y < \pi / 2 b^{\prime},
\\
-\pi / c^{\prime} \leq k_z < \pi / c^{\prime},
\end{array}
\right.
$$
and $u_{n{\bf k}}({\bf x})$ is commensurate with  the
'magnetic' unit cell. Here we assume the existence of a gap in the
quasiparticle spectrum around zero energy.

\section{3D spin Hall conductance and Chern number}

We define a spin current from the spin conservation law
\begin{equation}
\dot \rho_s({\bf x}) + \nabla \cdot {\bf j}_s^z=0,
\label{conservation}
\end{equation}
where
\begin{equation}
\rho_s^z({\bf x})=\frac{1}{2}\sum_{n\leq 0}
\int_{MBZ} \frac{d^3k}{(2 \pi)^3} \Phi_{n{\bf k}}^{\dagger}({\bf x})
\Phi_{n{\bf k}}({\bf x})
\label{spin-density}
\end{equation}
is the density of the $z$-components of the spins.  The conservation is
assured by the spin rotational symmetry around the $z$-axis in the
$^3$He-A phase Eq.(\ref{d-vector}). Note that, in the
particle-hole representation, the spin density is written by the sum
of $\Phi_{n {\bf k}}^{\dagger}({\bf x})\Phi_{n {\bf k}}({\bf x})$
which has the same form as the charge density of a state $(n, {\bf
k})$ in the usual representation. From Eqs. ({\ref{conservation})
and (\ref{spin-density}), the explicit form of the spin current for
the $z$ component is given by
 $$
 {\bf J}_s^z({\bf x})=\frac{1}{2}\sum_{n\leq 0}
\int_{MBZ} \frac{d^3k}{(2 \pi)^3} \Phi_{n{\bf k}}^{\dagger}({\bf x})
\frac{1}{i}\left[{\bf x},{\cal {H}}_U\right] \Phi_{n{\bf k}}({\bf
x}),
 $$
 which is also of the same form as the charge current in the usual
 representation.

 To drive the spin current, we apply an inhomogeneous magnetic field
 in the $z$-direction, ${\bf B}=B_z {\bf e}_z$, with a constant
 gradient, i.e. $\nabla B_z = const$. Then to the Hamiltonian
 (\ref{Hu}) the Zeeman coupling term ${\cal {H}}_U$ should be added,
 which in the particle-hole representation reads
 \begin{equation}
 {\cal {H}}_I=\frac{\mu_{\rm He}}{2}{\bf x} \cdot {\nabla} B_z.
 \end{equation}
This has the same form as the interaction between the charge
density and a constant electric field in the usual representation.
Then, one can see a mapping between physical quantities in the
Hamiltionian for Bloch electrons in the quantum Hall (QH) system
\cite{TKNN,Kohmoto,K-H-W} and that for superfluid $^3$He in the
particle-hole representation. See Table \ref{table2}.
\begin{table}
\caption{The mapping between physical quantities in the Hamiltonian
for Bloch electrons in QH system \cite{TKNN,Kohmoto,K-H-W} and that
for superfluid $^3$He in the particle-hole representation}
\begin{center}
\begin{tabular}{c|c}
Bloch electrons &  superfluid $^3$He
\\ \hline
e& $ 1/2 $ \\
$\rho_{e}$& $\rho_s^z$ \\
~ ${\bf J}_{e}$ & ${\bf J}_s^z$ \\
${\bf E}$& ${\bf \nabla} B_z$
\end{tabular}
\end{center}
\label{table2}
\end{table}

The mapping tells us that the expectation value of the spin current
obtained by the Kubo formula \cite{Kubo} has the same form as that
for the charged Hall current of 3D Bloch electrons\cite{K-H-W};
namely,
$$
<J_{si}^z>=\sigma_{ij}^s {\nabla}_j B_z  ,
$$
with a conductivity
\begin{eqnarray}
\sigma_{ij}^s&=&\mu_{\rm He} \sum_{n\leq 0,m} \int_{MBZ} \frac{d^3 k}{(2 \pi)^3}
\frac{<u_{n{\bf k}}|J_{si}^z|u_{m{\bf k}}><u_{m{\bf k}}|J_{sj}^z|u_{n{\bf k}}>
- (n \leftrightarrow m)}{(E_{m{\bf k}} - E_{n{\bf k}})^2}
\nonumber\\
&=&\frac{\mu_{\rm He}}{4}\sum_{n\leq 0} \int_{MBZ} \frac{d^3 k}{(2 \pi)^3 i}
\epsilon_{ijk} \left[{\bf \nabla}_{{\bf k}} \times {\bf A}_n({\bf
k}) \right]_k, \label{SH}
\end{eqnarray}
where ${\bf \nabla}_{\bf k}=\partial / \partial {\bf k}$ and ${\bf
A}({\bf k})$ is a gauge connection in ${\bf k}$-space
$$
{\bf A}_n({\bf k})=<u_{n{\bf k}}|\nabla_{\bf k}|u_{n{\bf k}}>,
$$
which was first introduced in this context by one of the authors \cite{Kohmoto}.

Following Ref. \cite{K-H-W}, we can relate the spin Hall conductance
(\ref{SH}) to the topological Chern number \cite{TKNN,Kohmoto} as
follows. The crystal momentum is parametrized as
\begin{equation}
{\bf k}=f_{a^{\prime}} {\bf G}_{a^{\prime}} + f_{b^{\prime}} {\bf
G}_{b^{\prime}}/2 +f_{c^{\prime}} {\bf G}_{c^{\prime}} \label{k}
\end{equation}
with $0< f_{a^{\prime}} , f_{b^{\prime}}, f_{c^{\prime}} \leq 1$,
and
$$
\int_{MBZ}d^3k = \int_0^1 df_c^{\prime} \frac{{\bf G}_{c^{\prime}}
\cdot{\bf c}^{\prime}}{c^{\prime}} \int_{T^2(f_{c^{\prime}})} d^2k
=\frac{2 \pi}{c^{\prime}}\int_0^1 df_{c^{\prime}}
\int_{T^2(f_{c^{\prime}})}  d^2k,
$$
where $T^2(f_{c^{\prime}})$ is a two dimensional torus formed by the
${\bf k}$'s with fixed $f_{c^{\prime}}$. Similarly,
$$
\int_{MBZ} d^3 k =\frac{2 \pi}{a^{\prime}}\int_0^1
d f_{a^{\prime}}\int_{T^2(f_{a^{\prime}})} d^2k
=\frac{\pi}{b^{\prime}}\int_0^1 d f_{b^{\prime}}
\int_{T^2(f_{b^{\prime}})} d^2k.
$$
One may introduce a vector ${\bf D}$ that satisfies
\begin{equation}
\sigma_{ij}=\epsilon_{ijk} D_k.
\end{equation}
We write ${\bf D}$ in terms of the primitive vectors of the
reciprocal lattice as
\begin{equation}
 {\bf D}=\alpha {\bf G}_{a^{\prime}} 
+ \beta {\bf G}_{b^{\prime}}/2
 + \gamma {\bf G}_{c^{\prime}},
\end{equation}
and obtain
$$
\gamma=\frac{1}{2 \pi} {\bf c}^{\prime}\cdot {\bf D}
=-\frac{\mu_{\rm He}} {16\pi^2}\int_0^1 df_{c^{\prime}}
\sum_{n\leq 0} \int_{T^2(f_{c^{\prime}})}
\frac{d^2 k}{2 \pi i} \left[{\bf \nabla}_{\bf k}
\times {\bf A}_n({\bf k})\right]\cdot
\frac{{\bf c}^{\prime}}{|c^{\prime}|}.
$$
Therefore, the integral
\begin{equation}
\sum_{n\leq 0} \int_{T^2(f_{c^{\prime}})}
\frac{d^2 k}{2 \pi i}\left[{\bf \nabla}_{\bf k}
\times {\bf A}_n({\bf k})\right]\cdot
\frac{{\bf c}^{\prime}}{|c^{\prime}|}
\label{Chern}
\end{equation}
is the first Chern number\cite{TKNN,Kohmoto,K-H-W} on the torus in
momentum space. For the discussions of the relation between
Berry's phase \cite{Berry} and Eq. (\ref{Chern}), see Refs.
\cite{G-K,G-K-0}. Because of its topological nature, the integral
is an integer which depends on $f_{c^{\prime}}$ when the Fermi
level for the quasi-particles is in a gap between magnetic Bloch
bands. Then,
\begin{equation}
\gamma=-\frac{\mu_{\rm He}}{16 \pi^2} N_{c^{\prime}},
\end{equation}
with $N_{c^{\prime}}$ an integer. We can also show that
\begin{eqnarray}
\alpha&=&-\frac{\mu_{\rm He}}{16 \pi^2} N_{a^{\prime}},
\nonumber\\
\beta&=&-\frac{\mu_{\rm He}}{16 \pi^2} N_{b^{\prime}},
\end{eqnarray}
where $N_{a^{\prime}}$ and $N_{b^{\prime}}$ are integers. So
finally we have
\begin{equation}
{\bf D}=-\frac{\mu_{\rm He}}{16\pi^2} {\bf G},
\end{equation}
where ${\bf G}$ is a lattice vector reciprocal to the original
lattice formed by ${\bf a^\prime},2{\bf b^\prime},{\bf c^\prime}$.
Thus, the spin current is shown to be
\begin{equation}
 <{\bf J}_s^z>=\frac{\mu_{\rm He}}{16 \pi^2} {\bf G} \times {\bf {\nabla}} B_z.
 \end{equation}
This result is the spin analogy in the present system of the
charged Hall current in 3DQHE \cite{K-H-W} .

\section{Summary and Discussions}

In summary, superfluid $^3$He-A is a good candidate system to
investigate the SHE, since this system has no spin-orbit
couplings, and has a spin rotational symmetry which ensures the
spin current conservation. We have presented the 3D extension of
the quantized spin Hall effect in the vortex state of a rotating
$^3$He-A superfluid film \cite{G-K} in an inhomogeneous magnetic
field. We have demonstrated the one to one correspondence between
the SHE in $^3$He-A in the case at hand and 3DQHE of Bloch
electrons previously studied in ref. \cite{K-H-W}, and thus have
predicted the existence of the 3d QSHE in this system, when
suitable conditions are satisfied.

Here we summarize in more details the conditions that are needed
for the predicted QSHE. As we have mentioned, the $^3$He-A
superfluid has to be in the vortex phase. The vortices normally
form a regular triangular lattice with the primitive vectors ${\bf
a}=(a,0,0)$ and ${\bf b}=(a/2,\sqrt{3}a/2,0)$ in the plane
perpendicular to the angular velocity ${\bf \Omega}=
(0,0,\Omega)$. Then we apply an one-dimensional periodic
potential, as what happens in an optical lattice, along an axis
which is not in the plane of the vortex lattice (the plane formed
by vectors ${\bf a}$ and ${\bf b}$. In the simplest case, this
axis can coincide with the axis of rotation, i.e. that of ${\bf
\Omega}$. This periodic potential should be weak, not to spoil the
superfluidity and the vortex phase. Then to drive the spin current
we need an inhomogeneous, static magnetic field in a fixed
direction, coupled to the nuclear magnetic moment of the $^3$He
nuclei, with constant gradient in its magnitude. When these
conditions are satisfied, we have shown an exact analogy between
the spin current of the quasi-particles in the present system (in
the particle-hole representation) and the charge current of the
Bloch electrons in a three dimensional periodic potential. Based
on the knowledge of latter \cite{K-H-W} with a connection to the
Diophantine equations (see ref. \cite{TKNN} and the third paper in
Ref. \cite{K-H-W}), a three dimensional quantized spin Hall effect
with non-zero Hall conductances is generically expected, when the
Fermi level of the Bogoliubov quasi-particles just completely fill
a certain number of Bloch bands, opening a gap in the spectrum of
quasi-particles.

Since the vortex lattice is created by rotation, the vortex
structure near boundary could be rather complicated. How spin
accumulation would be affected by conditions at the boundary is an
important issue for experimental detection. Theoretically this is 
a separate issue from the bulk effect we have addressed in this 
paper, and is certainly worth further study.

Finally we emphasize that in the above we have mentioned the
$^3$He-A phase as a familiar example of a $p$-wave paired Fermi
system. Actually the above discussions are valid also for cold
Fermi atomic gases that allow $p$-wave pairing in a certain
parameter regime. This remark opens the door for more candidate
systems that may experimentally realize the predicted QSHE.

This work is supported by the visiting program of the Institute of
Solid State Physics (ISSP), University of Tokyo in 2005 and 2006.
JG is supported by Grant-in-Aid for Scientific Research for the
Japan Society for the Promotion of Science under Grant No.
16740226. YSW is supported in part by the US NSF through Grant No.
PHY-0407187, and he thanks the warm hospitality from the ISSP.


\begin{references}
\bibitem{note} Tuning and enhancement of $p$-wave interaction in 
a cold Fermi atomic gas has been achieved and observed in ref.
\cite{regal03} through Feshbach resonance.
\end{references}
\end{document}